\documentclass[aps, prl, superscriptaddress, preprint]{revtex4-1}

\usepackage{graphicx}  
\usepackage{dcolumn}   
\usepackage{bm}        
\usepackage{amsmath}
\usepackage{SIunits}
\usepackage{times}
\usepackage{float}

\begin{document}

\title{Mapping of Yu-Shiba-Rusinov states from an extended scatterer}

\author{Markus Etzkorn}
\email[Corresponding author; electronic address:\ ]{m.etzkorn@tu-bs.de}
\email[Present address:]{ Insitut f\"{u}r Angewandte Physik, TU Braunschweig.}
\affiliation{Max-Planck-Institut f\"ur Festk\"orperforschung, 70569 Stuttgart, Germany}
\author{\normalfont\textsuperscript{,$\dagger$}~~Matthias Eltschka}
\thanks{These two authors contributed equally.}
\affiliation{Max-Planck-Institut f\"ur Festk\"orperforschung, 70569 Stuttgart, Germany}
\author{Berthold J\"ack}
\affiliation{Max-Planck-Institut f\"ur Festk\"orperforschung, 70569 Stuttgart, Germany}
\author{Christian R. Ast}
\affiliation{Max-Planck-Institut f\"ur Festk\"orperforschung, 70569 Stuttgart, Germany}
\author{Klaus Kern}
\affiliation{Max-Planck-Institut f\"ur Festk\"orperforschung, 70569 Stuttgart, Germany}
\affiliation{Physique de la Mati{\`e}re Condens{\'e}e, Ecole Polytechnique F{\'e}d{\'e}rale de Lausanne, 1015 Lausanne, Switzerland}

\date{\today}

\begin{abstract}
We investigate the spatial evolution of Yu-Shiba-Rusinov (YSR) states resulting from the interaction between Copper phthalocyanine molecules and a superconducting Vanadium (100) surface with submolecular resolution. Each molecule creates several YSR states at different energies showing distinctly different spatial intensity patterns. Surprisingly, on the molecules the largest YSR intensities are found not at the metal center, but close to one of the pyrrolic N-atoms, demonstrating strong molecular substrate interactions via the organic ligands. Energy resolved YSR maps reveal that the different YSR states originate from different molecular orbitals. We also follow the YSR states well-beyond the extent of the molecules and find clear oscillations of the YSR intensities without strong particle hole scattering phase differences. This is in contrast to expectations from a point scattering model and recent experimental findings on atomic impurities on superconductors. Our results can be explained by treating the molecular system as extended scatterer. Our findings provide new insights that are crucial to interpret the effects of a variety of magnetic systems on superconductors, in particular also those discussed in the context of Majorana bound states, which because of their size can not be considered point like as well.
\end{abstract}

\maketitle



The interaction with magnetic material can be used to strongly modify the properties of a superconductor. This has recently gained much attention in the context of Majorana bound states that can be induced by magnetic atomic chains or islands on superconductors~\cite{perge_2014, Ruby_2014-115, menard_2016, Feldman17}. On the atomic level, single magnetic impurities create resonances 
inside the superconducting gap, so called Yu-Shiba-Rusionv (YSR)~states~\cite{yu_1965, shiba_1968, rusinov_1969, balatsky_2006}. These are thought to be well-protected from the hosts electronic excitation spectrum and for this reason have been suggest to be used in quantum information storage. This however relays on the creation of a single pair of YSR states well protected from interactions with other YSR states or quasiparticle excitations.
YSR states have long been studied on the local scale by STM~\cite{yazdani_1997} and, in addition, have been employed as a probe for the symmetry of the order parameter~\cite{yazdani_1999}, to investigate competing Kondo-screening~\cite{franke_2011}, or scattering with different orbital channels~\cite{Ji_08}.
Two studies have also shown the impact of the multiplet structure of the impurity on the number of created YSR states and their observed scattering pattern~\cite{ruby_2016, Choi_2016}.
The vast majority of publications on YSR states report strongly localized YSR states essentially observed on the scatterer itself. Only few studies have investigated systems that exhibit YSR wave functions spatially extended well beyond the scale of the scatterer~\cite{menard_2015,ruby_2016, Ruby_2018, Choi_2018}, i.e. on the order of a few nanometers.
The simplest scattering model for a point scatterer in a three dimensional system with an isotropic Fermi surface with wave vector $k_F$ predicts a spatially oscillating wavefunction $\propto \sin\left(k_F r -\delta^\pm\right)/(k_F r)$ of the YSR-state~\cite{balatsky_2006, yazdani_1997, hudson_2001, menard_2015} (see Fig. 1~a). Here $\delta^\pm$ denote the scattering phase for particle and hole like YSR-states, i.e. the states above and below the Fermi level. The Fermi surfaces of the superconductor investigated experimentally so far (NbSe$_2$ and Pb), are rather complex and very anisotropic yielding a focusing effect of the YSR state intensities along specific spatial directions. Despite the rather complex shape of the Fermi surface the observed scattering pattern along those focusing directions can be surprisingly well described by the above formula~\cite{menard_2015,ruby_2016}. In particular, the observed phase difference between the particle and hole wavefunctions corresponds to the observed energies of the YSR states in agreement with the simple model. In the two experimental studies discussing the spatial evolution of the YSR states, the scattering impurity was a single $3d$-transition metal atom impurity~\cite{menard_2015,ruby_2016}, for which the central assumption of a point scattering potential ($i.e.$ the spatial extent of the potential to be much smaller than the Fermi wavelength $\lambda_F$) seems reasonable. However, in most metals $\lambda_{F}$ is of the order of~1\,nm. It, therefore, seems likely that the scattering potential created by an impurity, in particular when larger than a single atom, can extent beyond such size. 
In fact the recent studies of Majorana bound states use large structures with strong internal interactions~\cite{perge_2014, menard_2016} to create these in-gap states. Once not point like, a scattering potential can also be expected to show local variations; for example caused by non equivalent atom positions in the Fe chains on Pb(110)~\cite{Feldman17}. Even in the case of a single molecule of moderate size its interaction with the superconductor should lead to a scattering potential that easily extends over the length scale of $\lambda_{F}$.

\begin{figure}[ht]
     \centering	 \includegraphics[width=0.8\textwidth]{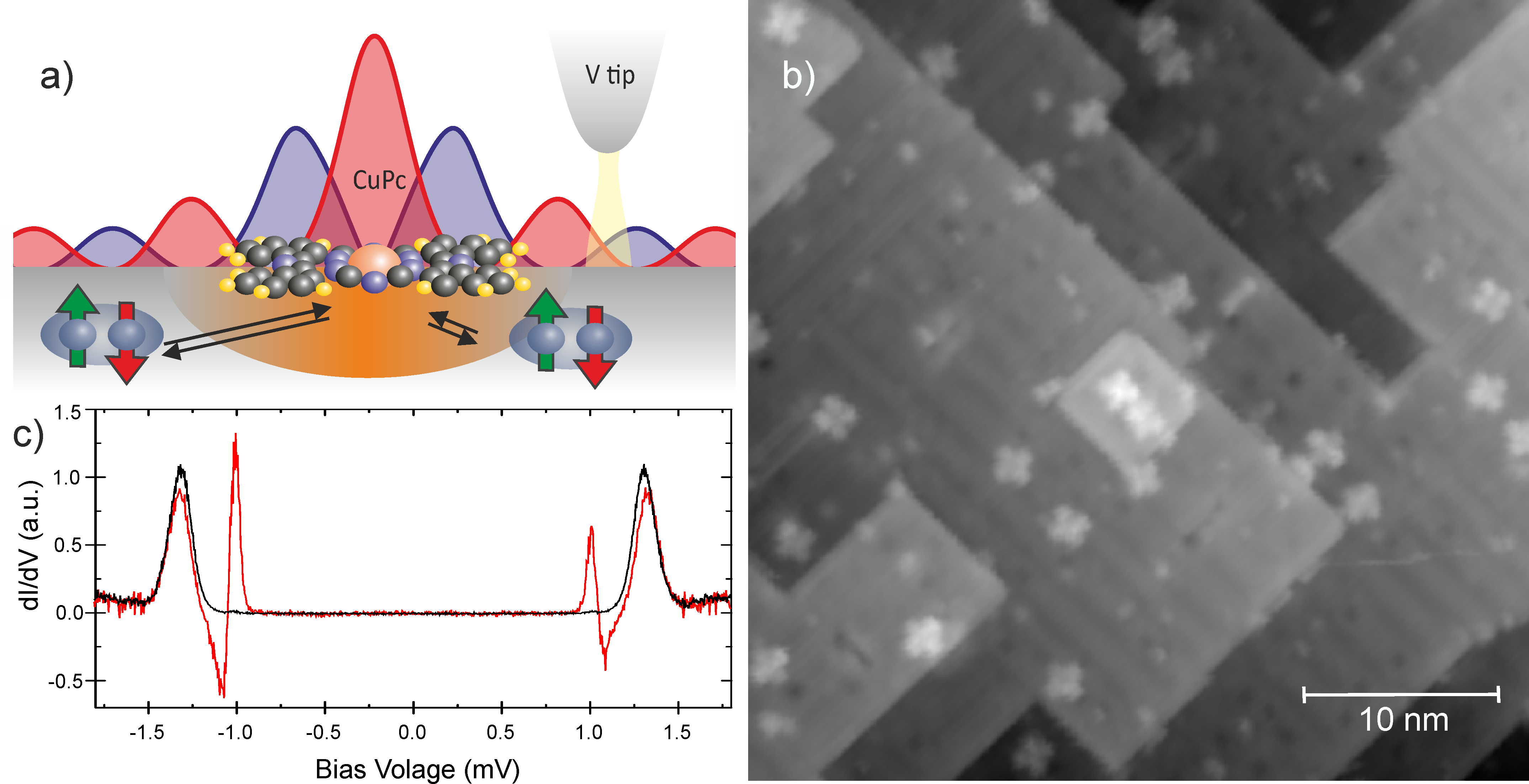}
	\caption{CuPc molecule on the V(100) surface. (a) The scheme illustrats the local interaction of the magnetic moment of the CuPc molecule with the superconducting state in the V(100) surface. The intensity of the resulting YSR bound states decay with distance while showing periodic oscillations. (b) The STM image ($I_\mathrm{T}=-50$\,\text{pA} and $V_\mathrm{T}=-8.5$\,\text{mV}, 36\,nm$\times$36\,nm) shows various terraces of the V(100) surface. The faint black lines perpendicular to the steps result from the $(5\times1$) reconstruction. CuPc molecules show a fourfold cloverleaf-shape due to the Pc rings around the Cu ion. c) Conductance spectrum in the center of one molecule that shows the most commonly observed spectral features (red). The spectrum shown in black was measured far away from the molecule. The superconducting gap of sample and tip are $760$ and $570\, \mathrm{\mu eV}$, respectively. (Stabilization conditions: $V_T= -2.2\, \text{mV}, I_T = 150\,\text{pA}$)}
	\label{fig_1}
\end{figure}

Here, we investigate the YSR states locally induced by a scattering potential that can not be considered pointlike, but that is extended on the scale of $\lambda_{F}$. We use metal-organic copper phthalocyanine (CuPc) molecules as a scatterer adsorbed on a superconducting V(100) substrate. Metal-Pc molecules are well studied almost planar molecules that adsorb flat on surfaces. They are known for their strong interaction between the molecule and metallic surfaces not only through the metallic cental atom, but also through the ligands, in particular the adjacent pyrrolic N atoms~\cite{ste10, Lach12}. This makes them prime candidates for the creation of an extended complex scattering potential for Cooper pairs. CuPc has a Cu$^{2+}$ $d^{9}$-ground state with a resulting spin and orbital magnetic moment of $S = 1/2, L=0$ in powder~\cite{mugarza_2012, Lee_87, Harrison_64}. Density functional calculations find the unpaired electron in an $b_{1g}$ orbital with mixed Cu $3d_{xy}$ and N $2p$ character, while the highest occupied and lowest unoccupied molecular orbital HOMO (LUMO) is an $a_{1u}$ ($e_{g}$) orbital localized exclusively on the organic backbone~\cite{Rosa_94, Marom_2011}. Despite the strong interactions and charge transfer between the molecule and several metallic substrates, experiments have shown that the spin 1/2 configuration of the Cu center remains unchanged on many surfaces~\cite{ste10, Gargiani2013, Yoshizawa2017}. The significant charge transfer of approximately one electron even on noble metal surfaces adds an additional magnetic moment on the ligand by a partial filling of the LUMO state~\cite{ste10, mugarza_2011, mugarza_2012, Yoshizawa2017}, yielding multiple, spatially extended magnetic moments on one molecule and, thus, a spatially-varying and not point like scattering potential. We deposited these molecules on a $(5 \times 1)$ oxygen reconstructed V(100) superconductor, where they create multiple YSR states with a complex pattern in real space. In the following, we investigate their spatial evolution and energies by using scanning tunneling microscopy at ultra low temperatures of $15\,$mK.

In Fig.~\ref{fig_1}(b), an STM image of the V(100) surface measured with a superconducting V-tip shows various terraces, with dark lines indicating the $(5\times1)$ reconstruction of the V(100) surface~\cite{koller_2001, kralj_2003} and fourfold clover-shaped CuPc molecules adsorbed on the surface. The molecules are found in several adsorption conformations differing in their position on the reconstruction and the orientation of the molecular symmetry axes with respect to the substrate atomic lattice. The different conformations strongly influence the YSR states and variations in the number of (resolved) YSR states, their energies as well as the spectral intensities are observed (see Fig. 1 in SI). In a large fraction of molecules, we find the YSR states to be rather intense, often larger than the quasi-particle coherence peaks, and with energies of $E_{i} \approx \Delta_{\text{sample}}/2$ (with $\Delta$ being the energy of the quasiparticle excitation gap). One representative example is shown in Fig.~1 c). Note that the spectrum is measured with a superconducting tip and, therefore, is a convolution of the sample density of states with the superconducting density of states of the tip. Thus, the coherence peaks are located at $\pm(\Delta_{\text{tip}}+\Delta_{\text{sample}})$ and the YSR states at $\pm(\Delta_{\text{tip}}+E_{0})$. As can be seen from a comparison between a spectrum on the molecule and on the V surface (Fig.~1 c)), the convolution with the superconducting density of states (DOS) of the tip results in the large negative differential conductance on the high energy tail of the YSR states that affect the shape of the quasiparticle peaks. Nevertheless, the size of the quasiparticle excitation gap of the V-sample remains unchanged in the presence of the molecule. In the following, we will focus on a detailed investigation of the YSR states from molecules with the most common adsorbtion confirmation that show YSR spectra similar to the one presented in Fig.~1c). They are adsorbed on a low symmetry site of the reconstruction, with the long axis of the molecule rotated by about 10$^{\circ}$ from the crystallographic [100] direction. We find YSR states showing intensity modulations not only within the extent of the molecule, but also in its vicinity, where the YSR intensities decay within the distance $x\approx \pm 30\,\text{\r{A}}$ away from the center of the molecule by more than three orders of magnitude.

\begin{figure}[H]
	\centering
     \includegraphics[width=0.8\textwidth]{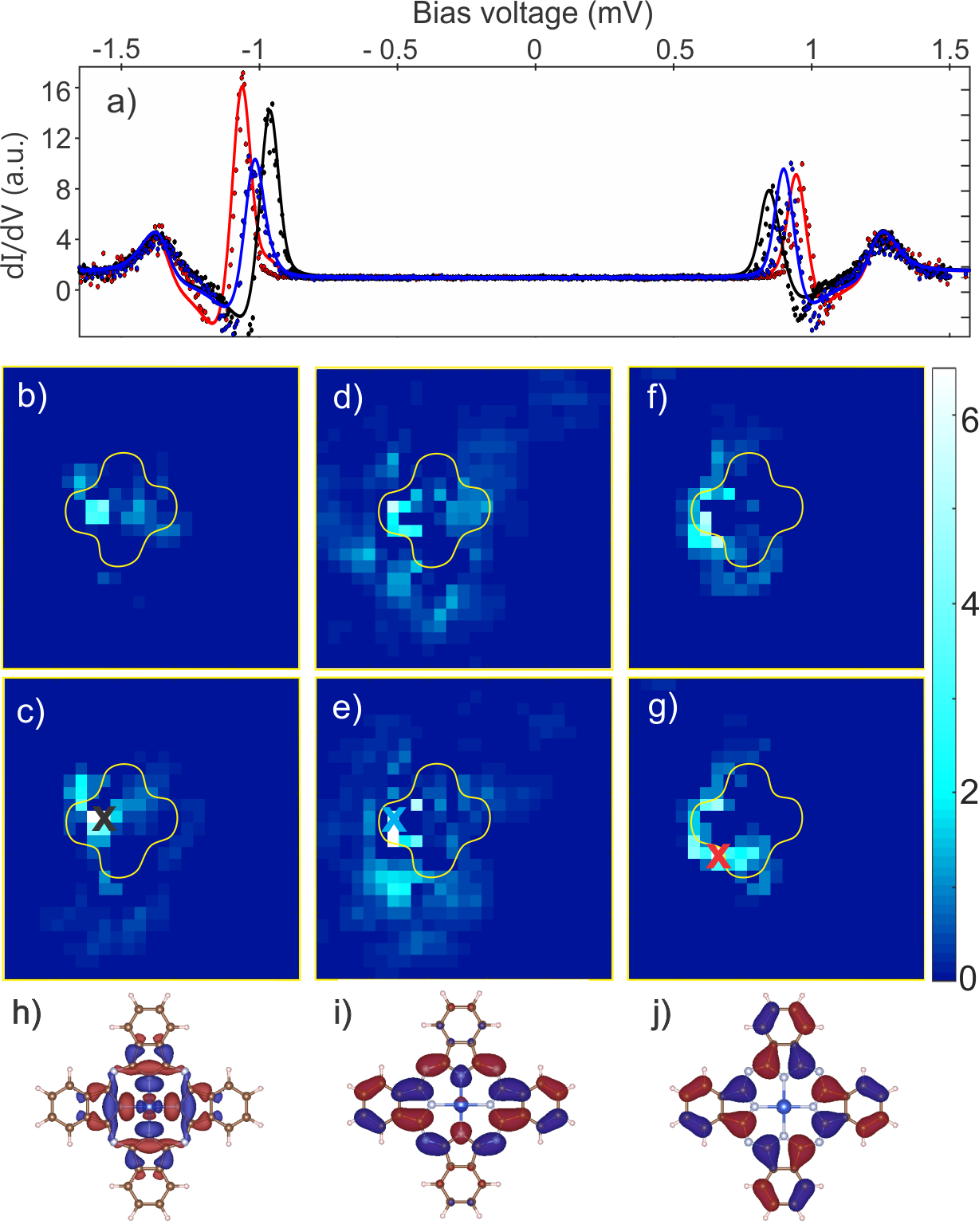}
	\caption{a) three dI/dV-spectra measured on different positions on the molecule that show important contributions of three YSR states (black $E_{1} = 355\, \mathrm{\mu eV}$, blue $E_{2} = 420\, \mathrm{\mu eV}$ and red $E_{3} = 445\, \mathrm{\mu eV}$). The measuring positions are marked by crosses in c), e) and g). The red lines are fits to the spectra (see text for details). b)-g) two dimensional distribution of the intensities of these three YSR states for particle (upper row) and hole (lower row) like states. b)-c) for the state at $E_{1} = \pm 355\, \mathrm{\mu eV}$, d)-e) $E_{2} = \pm 420\, \mathrm{\mu eV}$, f)-g) $E_{3} = \pm 445\, \mathrm{\mu eV}$. The molecular orbitals are results of DFT calculations for the gas phase for SOMO $b_{1g}$ (h), LUMO $e_{g}$ (i) and HOMO $a_{1u}$ orbitals.}
	\label{fig_2Dmap}
\end{figure}

To investigate the YSR states and their spatial dependence in more detail, we made a spectroscopic map of 25 $\times$ 25 spectra over one CuPc molecule and its close vicinity. As can be seen from Fig.~\ref{fig_2Dmap} a) the molecule shows three different YSR states at $E_{1} = 355\, \mathrm{\mu eV}, E_{2} = 420\, \mathrm{\mu eV}$ and $ E_{3} = 445\, \mathrm{\mu eV}$, respectively. 
The solid lines in the figure are fits to the spectra. For the fits, we use the quasiparticle density of states of the superconducting tip and sample and model the YSR states by Loretzian peaks at energies $\pm E_{i}$ in the sample density of states (see SI for details). We find a very good overall agreement between the fits and the 625 spectra using the intensities of the three YSR states as the only variable fit parameters. Fig. 2b)-g) show the resulting intensity distribution for each YSR peak which reveal the strong spatial dependence of the YSR wave functions. The effect of the low symmetry adsorption site on the $(5 \times 1)$ V-reconstruction is reflected by the variation of the YSR intensities that do not follow the original fourfold symmetry of the molecule. We find that the maximum of the wave functions of the YSR state with the lowest energy (i.e. the strongest scattering potential) is located very close to one of the pyrrolic N atoms. For the YSR states with higher energies the maximum YSR intensity shifts further away from the central metal atom onto the ligands. The spectral weight of the YSR states center of mass is clearly shifted towards one pyrrolic N-atom of the ligand which suggests that the strongest interaction between the molecule and the superconductor is located there. Nevertheless, each YSR state has some spectral weight on other organic side groups of the molecule. This experimental observation already indicates a molecule-substrate interaction which is significantly extended over the size of the molecule, and therefore over sizes comparable to $\lambda_{F}$, creating a spatially varying scattering potential for the superconducting condensate. This documents the importance of the ligand substrate interaction, that has been previously found in DFT studies on magnetic substrates~\cite{Lach12}.

Recent publications have investigated the effect of the orbital structure of atomic impurities to explain the origin of multiple YSR states and their spatial intensity variations~\cite{ruby_2016, Choi_2016, Ruby_2018, Choi_2018}.
They found that the YSR intensities probed by STM over the impurity are determined by the shape of the atomic orbital that creates the YSR state via its interaction to the superconductor. Translating this idea to a molecule, we can also expect different molecular orbitals to create YSR states at different energies. However, it is important to recall that YSR-states can only result from the interaction of a partially filled impurity state with the superconducting Cooper pairs. In the gas phase only the $b_{1g}$ molecular orbital located on the metal center and the neighboring N-atoms is singly occupied. It is observed in many systems that upon adsorption the charge transfer from the substrate to the molecule induces an additional unpaired electron into the CuPc LUMO state~\cite{ste10, mugarza_2011, mugarza_2012, Yoshizawa2017}. In our case, considering the symmetry breaking due to the substrate, the molecule the intensity profile for the two YSR states with lowest energy and therefore strongest interaction conform to the gas phase SUMO and LUMO orbital shape rather well (see Fig. 2 b)-e)). Following this argument the intensity distribution of the YSR state with the weakest interaction (Fig. 2 f)-g)) is reminiscent of the HOMO orbital. In the gas phase, even when considering charge transfer the HOMO does not show any significant spin polarization~\cite{mugarza_2011, mugarza_2012}. However, we expect that the hybridization with the substrate induces a finite spin polarization into the HOMO as well. Alternatively, this state could also be explained by a lifted degeneracy of the LUMO$_{xz}$ and LUMO$_{yz}$ orbitals due to the low symmetry adsorption position. The split orbitals have the same shape (shown in Fig.~2~i) but are rotated by $90^{\circ}$ with respect to each other. For the reasons discussed above a strong spin polarization is expected in both orbitals due to charge transfer. The different molecular orbitals causing the spatial variations also give a natural explanation for the observed different YSR state energies, as each orbital will interact differently with the superconducting substrate. We want to note that the maximum energy difference between the YSR states is  $100 \, \mathrm{\mu eV}$ in the system investigated here. These states could be easily mistaken for one single state. This illustrates the importance of a careful choice of the YSR system, if applications in quantum information processes are envisioned, as the mixing of different YSR states will lift the protection of the state.

In principle, multiple YSR states can also result from magnetic anisotropy~\cite{Zitko_2011, Hatter_2015}, coupling to bosonic excitations like vibrations, or different angular momentum scattering channels~\cite{Flatte_1997, Ji_08}. However, a magnetic anisotropy leads to a splitting of a single YSR peak and, therefore, the ratios of the YSR intensities must be constant over the molecule, which strongly disagrees with our data. For the same reason, we can exclude bosonic excitations as an explanation. The measured intensity distribution would in fact be in agreement with an interpretation of different angular momentum scattering channels causing the YSR peaks~\cite{Flatte_1997}. However, in general it can be expected that different angular momentum channels have strongly different YSR energies. Therefore, it is unlikely that three YSR states within an energy window of less than $100 \, \mathrm{\mu eV}$ are caused by different angular momentum channels.
We can, therefore, identify the YSR states on the molecule  to result from different  molecular orbitals. As there is a strong interaction between the molecule, in particular its ligands, to the substrate, we can directly infer from these measurements that the scattering potential caused by the molecule will extend at least over the size of the molecule and can not be considered point like on the scale of the Fermi wavelength. This has direct consequences for the YSR intensities measured far away from the molecule.

\begin{figure}[ht]
	\centering
      \includegraphics[width=0.8\textwidth]{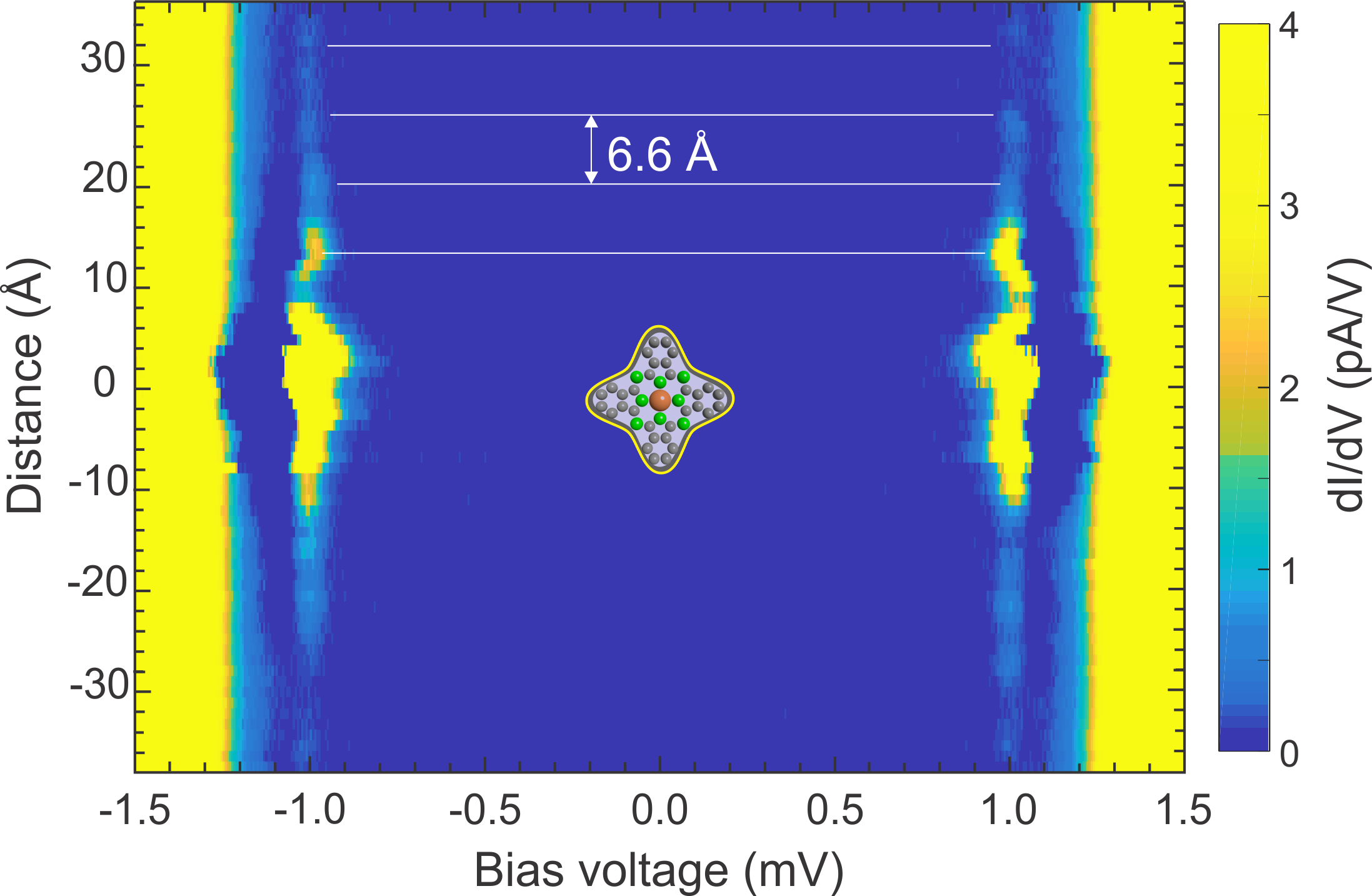}
	\caption{a) Differential conductance measurements along a line crossing the molecule (inserted to show its position and size). The $dI/dV$ spectra feature peaks at $\Delta_{sample}+\Delta_{tip}\approx \pm 1.35$\,meV due to the tunneling between the coherence peaks of the superconducting quasi-particle DOS of tip and sample. The YSR state occurs in the gap at energies ($ E+\Delta_{tip}\approx\pm 1$\,meV) and decays from the center of the CuPc molecule while oscillating in both intensity and energy. The white lines are equidistance and show the oscillation period. At each point, the tunnel contact is stabilized at $I_\mathrm{S} = -150$\,pA and $V_\mathrm{S}=-2$\,mV. 
}
	\label{fig_2}
\end{figure}

We find strong variations of the YSR wave functions well beyond the extent of the molecule, in particular for the YSR state at intermediate energy (420$\, \mathrm{\mu eV}$, see Fig.~2~d)-e)). The states show intensity modulations in both particle and hole states. To investigate the spatial dependence of the YSR state well beyond the molecular scale in more detail, we acquired 75 $dI/dV$ spectra on a line crossing a CuPc molecule. The results are shown in Fig.~\ref{fig_2} as false color plot of the spectral intensity as function of position and energy. The vertical features at $\approx \pm 1.3$\,mV result from the quasiparticle coherence peaks. More importantly, strong YSR states are observed within the gap at $\approx \pm 1$\,meV corresponding to particle and hole states that show a rich spatial dependence. Although the intensities of the quasi-particle excitations strongly decrease with increasing distance $x$ from the CuPc structure, they are detectable up to $x\approx \pm 30$\,\r{A} from the center of the molecule. In addition, the YSR intensities show oscillations with a period of approximately 6.6\,{\AA} as indicated by the white lines in Fig.~3, which are approximately in phase for particle and hole like YSR states.



A closer look at Fig.~3 reveals that the energy of the YSR peaks modulate with position which clearly shows that several YSR states contribute to the spectrum, as found also in the first molecule. Again, we can rationalize the spectra in Fig. 3 by fitting two pairs of peaks having an energy of $E_1 = \pm 465\, \mathrm{\mu eV}$ and $E_2 = \pm 490\, \mathrm{\mu eV}$ with the intensities of the YSR states as the only varying parameters. This allows us to describe all the rich phenomena, i.e. the strong intensity and the apparent energy oscillations.

\begin{figure}[ht]
	\centering
      \includegraphics[width=0.6\textwidth]{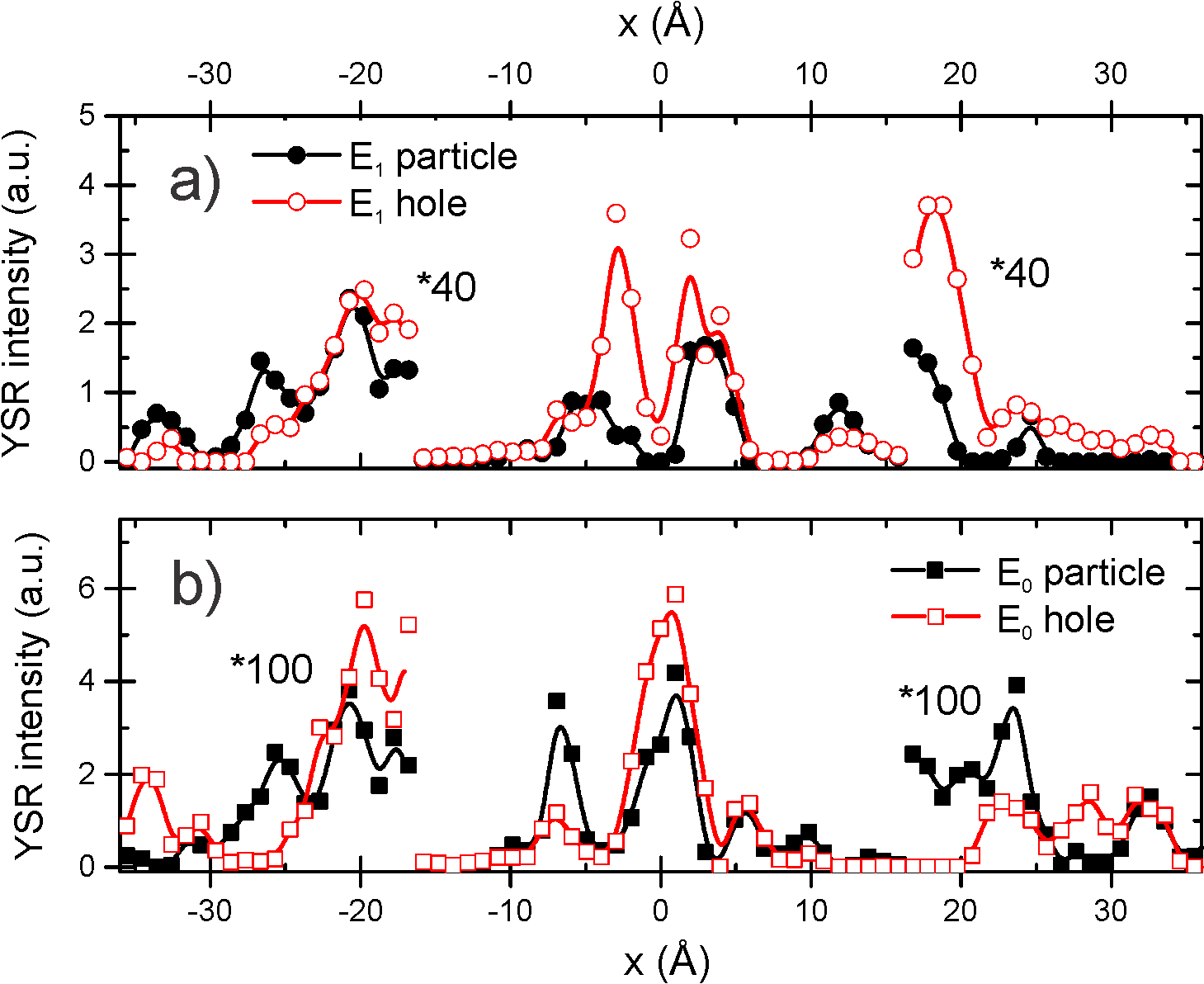}
	\caption{Spatially resolved intensities of the YSR states with particle and hole character for the peak at $E_1 = \pm 465\, \mathrm{\mu eV}$ (a) and $E_2 = \pm 490\, \mathrm{\mu eV}$ (b) energy as resulting from the fits of the measured spectra shown in Fig.~2. Oscillations of the YSR intensities occur that are out-of-phase concerning the two energies, but in phase for particle and hole contributions. The former effect explains the apparent energy oscillations of the YSR states seen in Fig.~2. 
}
	\label{fig_3}
\end{figure}

Figure~4 a) and b) reveal the rich structure of the spatial dependence of the intensities for the particle and hole contributions of the two YSR states, respectively, as extracted from the fitting. In particular, they show that the apparent oscillations of the YSR state energies with position can be simply explained by the out-of-phase intensity modulation of the two peaks having slightly different energies. This is well in agreement with our observations in the grid measurements shown in Fig.~2, reflecting different YSR states. The periodic YSR intensity variations are governed by a dominating Fermi wave vector, despite the complex Fermi surfaces of V.

As shown in figures 2 to 4, the variations of the particle and hole contributions of the YSR states are approximately in phase, also well outside the region over which the molecule extends. Assuming weak coupling, the normalized energies $\frac{E_i}{\Delta}$ of the YSR states range from 0.48-0.63. For a simple point scattering model we would, therefore, expect a large phase difference where the particle and hole wave function variations are out-of-phase~\cite{rusinov_1969,balatsky_2006} (see SI for details). This is clearly not the case in our measurements. We, thus, find two significant deviations from the simple point scattering model: the intensity variations are more complicated than sinusoidal and the expected sizable phase differences are not experimentally observed (see Fig.~4). The simple model assumes an isotropic Fermi surface which is certainly not the case for V. However, we recall that the other experimental investigations that followed the YSR states over regions of several nm studied systems with highly anisotropic and complex Fermi surfaces as well, and both find the oscillations and the phase shift of the YSR states to be in accordance with this simple model~\cite{menard_2015,ruby_2016}. The most obvious difference is that in these two studies single atomic impurities were used, where the approximation of a point scattering potential is reasonable. In contrast, here we investigate a molecular system causing an extended scattering potential. At sufficiently large distances, the scattering potential can be regarded point like again, however, the fast decay of the YSR states intensities limits the detection radius. In the range, in which we are able to follow the YSR wave functions, the extended and spatially inhomogeneous scattering potential will lead to an averaging of phase shifts obtained from scattering at different locations and a resulting strong reduction of the average phase shift.

In the next step, we address the tunneling mechanism of particles into the YSR states as well as the life time of excited quasiparticles in YSR states. In a recent study, it was shown that both single and multiple tunneling processes into YSR states occur, each of which resulting in a different tunneling spectrum~\cite{Ruby_2014-Aug}. The relative contributions strongly depend on the relaxation rate of the excited YSR state. The longer their lifetime the earlier multiple tunneling processes become important. The observation of a strong negative differential conductance at the high energy shoulder of YSR states is a strong  experimental indication of dominating single particle tunneling contributions in our spectra~\cite{Ruby_2014-Aug}. This finding in return assures that the measured intensity variations reflect the YSR density of states directly. A true BCS like quasiparticle density of states with a strictly empty gap requires a thermally driven  phonon annihilation process to decay an excited YSR state. Due to the low temperature at which we performed the experiments $(15 \,$mK) such processes are essentially quenched. However, we find the V quasiparticle density of states to clearly deviate from the BCS like form, leaving a finite number of generic quasiparticle excitations below $\Delta$. These states will provide efficient direct decay channels for the YSR states, explaining the limited lifetime of the YSR states even at very low temperatures. At the same time, the fact that the YSR peaks in the spectra result from single particle tunneling implies that the YSR state is depopulated before the next tunneling event. This sets a lower bound to the YSR life time broadening. The width of the YSR states is determined by the energy resolution of our experiment, which is quantum limited due to effects of the finite capacitance of the STM junction~\cite{Ast16}. This sets an upper bound for the life time broadening of these YSR states. Our data is fully compatible with YSR peaks that result from single particle tunneling in which the life time broadening of the YSR states of about $10\,\mathrm{\mu eV}$ (see SI for details). 

In passing, we note that the measurements in Fig.~2 and 3 show spatially oscillating YSR state intensities at different energies. These experimental observations directly exclude any sizable effect caused by the presence of the STM tip, e.g. due to the electric field or mechanical deformations of the molecule, as these would result in a monotonous modification of the YSR states when moving the tip away from the molecule.

In summary, we have investigated the YSR wave functions created by CuPc molecules on a $(5 \times 1)$ reconstructed V(100) surface. The most commonly observed molecular adsorption conformation is a low symmetry adsorption site, and we have investigated the YSR states of such molecules in detail. Each of these molecules show several YSR states having similar energies of $\frac{E_{i}}{\Delta} \approx 0.5$. We find rich local variations of the YSR intensities that strongly differ for states with different energy, which we follow with submolecular resolution. We can explain the local variations measured over the molecule with the creation of YSR states at different energies due to the different interaction of the superconductor to different molecular orbitals. The spatial distribution of each YSR state on the molecule reflects the form of the molecular orbital it is created from, in close resemblance to recent measurements on atomic impurities~\cite{ruby_2016, Choi_2016}. The center of mass of the spectral intensities is located close to a pyrrolic N atom on one ligand of the CuPc, demonstrating strong interactions between the molecule and the superconducting condensate from the organic backbone. For each molecule, the YSR states with lowest energy (i.e. highest binding energy) are located towards that N atom. The YSR state intensities can be followed well beyond the molecule itself. The modulations show little phase difference between particle and hole contributions which we attribute to be a direct consequence of the extended scatterer, 
contrasting experimental results for single atom impurities~\cite{menard_2015,ruby_2016}. 
We have discussed the impact of a finite density of states in the sample gap, deviating from the strict BCS form. These states provide efficient decay channels for YSR states and have to be taken into consideration when interpreting STM measurements on YSR states in particular for experiments performed at ultralow temperature.
The observed effect of extended scatterers creating a structured scattering potential on the scale of $\lambda_F$ is likely to be a rather common occurrence and is expected to appear in a variety of systems. The resulting spatial dependence of the YSR states is important also when considering the interplay of several scatterers for example in the context of Shiba bands or on Majorana bound states~\cite{perge_2014, menard_2016}.

\section{Methodes}
For our measurements, we used an STM operating in ultrahigh vacuum (UHV) at a base temperature of $15$\,mK~\cite{assig_2013}.  The V(100) crystal was cleaned by repeated Ar sputtering (at $1\,$kV) and consecutive annealing to $\approx 700\, $K  to form the (5x1) oxygen reconstruction from surface near O-contaminations~\cite{koller_2001, kralj_2003}. Afterwards small amounts of CuPc molecules ($\approx$ 1~\% of a monolayer) were deposited with organic molecular epitaxy using a cell at $\approx 660\,$K with the sample termperature held at $300\,$K. The V-tip has been made from V wire (purity 99.8~\%) that was cleaned by field emission on a clean V(100) crystal, prior to the experiment. The details of the tip preparation are discussed elselwhere~\cite{eltschka_2015}.
The details of the data analysis and in particular the procedure how to take into account the energy resolution broadening due to the finite capacitance on the tunnel contact and how to determine the junction parameters is shortly described in the SI and in much greater detail in~\cite{Ast16}.

\section{Acknowledgment}
It is our pleasure to acknowledge contributions from Rico Gutzler, for providing the DFT calculations of the CuPc orbital structure as well as helpful discussions with Alexander Balatsky, Peter Oppeneer, and Thomas Jung.

\section{Author contributions}
M. Et., M. El., B. J., C. A., and K. K. designed the experiment. M. El. and B. J. performed the measurements with the help of M. Et. M. Et., C. A. and M. El. analyzed the data. All authors discussed the analysis and results. M. Et. and M. El. wrote the manuscribt with inputs from all authors.

\newpage
\bibliographystyle{mystyle}
\bibliography{ms_bib}

\end{document}